# DISCOVERING THE CRITICAL NUMBER OF RESPONDENTS TO VALIDATE AN ITEM IN A QUESTIONNAIRE: THE BINOMIAL CUT-LEVEL CONTENT VALIDITY PROPOSAL

**Helder Gomes Costa**
Universidade Federal Fluminense
Rua Passos da Pátria, 156
Niterói, RJ , Brazil, CEP 24.220-045
heldergc@id.uff.br [Corresponding author]

**Eduardo Shimoda**
Universidade Candido Mendes
Rua Anita Peçanha, 100
Campos dos Goytacazes, RJ, 28030-335, Brazil
shimoda@ucam-campos.br

**José Fabiano da Serra Costa**
Universidade do Estado do Rio de Janeiro
Rua São Francisco Xavier, 524
Rio de Janeiro, RJ, 20550-900, Brazil
fabiano@ime.uerj.br

**Aldo Shimoya**
Universidade Candido Mendes
Rua Anita Peçanha, 100
Campos dos Goytacazes, RJ, 28030-335, Brazil
aldoshimoya@yahoo.com.br

**Edilvando Pereira Eufrazio**
Universidade Federal Fluminense
Rua Passos da Pátria, 156
Niterói, RJ , Brazil, CEP 24.220-045
Intituto Nacional de Tecnologia
Av. Venezuela, 82
Rio de Janeiro, RJ, 20081-312, Brazil
edilvandopereira@id.uff.br, edilvandopereira@id.uff.br l

## ABSTRACT

The question that drives this research is: "How to discover the number of respondents that are necessary to validate items of a questionnaire as actually essential to reach the questionnaire's proposal?" Among the efforts in this subject, Lawshe [1975], Wilson et al. [2012], Ayre and Scally [2014] approached this issue by proposing and refining the Content Validation Ratio (CVR) that looks to identify items that are actually essentials. Despite their contribution, these studies do not check if an item validated as "essential" should be also validated as "not essential" by the same sample, which should be a paradox. Another issue is the assignment a probability equal a 50% to a item be randomly checked by a respondent as essential, despite an evaluator has three options to choose. Our proposal faces these issues, making it possible to verify if a paradoxical situation occurs, and being more precise in recommending whether an item should either be retained or discarded from a questionnaire.

***Keywords*** Binomial distribution, Content Validation, Content Validity, CVR, CVI, Item validation, Item validity, Questionnaire, Survey.

# 1 Introduction

The discussion over the validity of the components that constitute research questionnaires is not novel. Over a century ago, Riley [1904] identified the primary challenges associated with utilizing questionnaires for indirect questions and advocated for direct oral inquiries as a more effective method of obtaining information from respondents, given the respondent sample is limited.

Since the publication of Riley [1904], extensive debate over the validity of questionnaires as a research instrument has occurred. A search conducted in October 2024 on Web of Sciences Core Collections using the query *"questionnair\* (Topic) and valid\*(Topic)"* returned 285,542 non-retracted documents or publications. The search yielded Riley [1904] as the earliest known work on the platform, which also emerged as the earliest from a analog search carried out in Scopus hrough the query "(TITLE-ABS-KEY (questionnar\* OR survey\*) AND TITLE-ABS-KEY (valid\*))", that resulted in 312,715 non-retracted records found.

Among subsequent studies concerning the development of scales for social questionnaires, we emphasize the following as seminal works:

- Thurstone [1928] devised a method for evaluating attitudes, emphasizing the continuous assessment derived from comparing respondents' agreement and disagreement with a series of assertions.
- Likert [1932] provided a methodology for constructing scales to assess attitudes. This approach emphasizes the necessity for all statements to reflect desired behavior rather than factual ones. Likert [1932] advocates for the implementation of symmetrical five-point scales.
- Miller [1956] investigated the optimal number of options that should be available in a attitude measurement scale, determining that such scales should offer $7^{+2}_{-2}$ choices to the respondents. In other words, as stated in Miller [1956], reactions are enhanced when the scale comprises 5 or 9 symmetric points.

In the subject of test validation, we highlight the contribution of Cronbach [1951], who introduced the statistical coefficient alpha to assess the inner workings of tests. The suggested coefficient gained prominence as Cronbach's alpha.

In another path of questionnaires validation, the work of Lawshe [1975] proposed the *Content Validity Ratio (CVR)* as a metric for validating the variables, aspects or questions that are essential to be part of an analysis. According to the *Content Validity Ratio (CVR)* values, Lawshe [1975] established the required minimum number of respondents needed, depending upon the sample size, for determining if an item is essential for measuring a property in a questionnaire. Subsequently, the calculus of CVR was reviwed by Wilson et al. [2012], a significant study that rectified the CVR values presented in Lawshe [1975]. Looking to better understand how CVR is computed, Ayre and Scally [2014] pointed out that the correction proposed in Wilson et al. [2012] is questionable, once it used a normal approximation instead of using a binomial modelling. Therefore, Ayre and Scally [2014] recalculated the minimum number of evaluators needed to agree with the statement that an item is "Essential". To do that, Ayre and Scally [2014] assumed the distribution as binomial, but maintained $p = 1/2$ as the hypothesized probability of agreeing an item as "Essential". To highlight the influence of these works another researches, we present Table 1 that illustrates the number of citations these works received from Scopus by October, 2024. We also mention that the importance of CVR is shift raised by the proposal of Lynn [1986] to the calculus of another metric: *Content Validity Index (CVI)* which has CVR in this core. This article had been cited by 4,006 other documents indexed in Scopus by October, 2024.

Table 1: A summary of citations received by articles that proposed and revised CVR computation

| Source | Number of citations |
|---|---|
| Lawshe [1975] | 4,005 |
| Wilson et al. [2012] | 382 |
| Ayre and Scally [2014] | 674 |

Therefore, to emphasize the importance of CVR in validation of social research and how it is spread around the world, we highlight here a piece of bibliometry about the works of Lawshe [1975], Wilson et al. [2012], Ayre and Scally [2014] and Lynn [1986]. First, a search performed on October, 10[th] 2024 was carried out in Scopus database using the query "REF (A quantitative approach to content validity OR Recalculation of the critical values for Lawshe's content validity ratio OR Determination and quantification of content validity)" resulting in finding non-retracted documents that mentioned at least one of these four previous works. The results from this search were exported into a bibtex file and analysed with the support ob Biblioshiny (Aria and Cuccurullo [2023]) with is a web based tool that runs the bibliometrix (Aria and Cuccurullo [2017]) package written in R (R Core Team [2022]) called in the RStudio IDE (RStudio Team [2020]).

Based on analysing the metadata of the records retried, we discovered that Lawshe [1975], Wilson et al. [2012], Ayre and Scally [2014] or Lynn [1986] were mentioned in $8,412$ non-retracted documents, spread in $2699$ different sources and were authored by $20,519$ different authors from 115 countries covering all the planet continents. We also claim that the number of publications on this subject is exponentially increasing along of almost 50 year.

But, despite the effort and the valuable results already reached in the calculus of the CVR, it still remains some features its computation and use should be improved. In subsection 3.4 we highlight a discussion about this point.

## 1.1 Objective

In present work we aim to propose the Binomial Cut-level Validation (BCV) that faces the remaining criticisms to CVR in validating if an item is essential in a questionnaire. In other words, the main goal of our proposal is to provide a more precise method to identify whether an item in a questionnaire is either essential or unnecessary, and even check if there is any kind of paradoxical perception about this question.

## 1.2 Highlights about the contribution

The main improvements of our proposal when comparing it against the previous ones are:

- It uses a binomial distribution assuming $p$ as a function of the number of options the evaluator has to choose in the form, instead of using a normal approximation and assuming $p = \frac{1}{2}$.
- It provides a kind of "double check validation" once it applies a double-tailed distribution. This double check allows to:
  - verify if an item is either "essential", or, "unnecessary", and,
  - check the presence of paradoxical inconsistencies in the results.

## 2 Recent researches using CVR

Researching in the topics addressed in the recent articles, we perceived an increasing use of CVR in topics such as Education, Healthcare, Quality of life, mainly with focus in children and women needs. As it follows, we mention a non-exhaustive sample of articles that should better exemplify this perception,

- Education and trainning
  - Aiming to discover the degree of knowledge sharing among faculty members, Sanjari and Soleimani [2024] proposed the Knowledge Sharing Behavior Scale (KSBS). In this research, the Lawshe's content validity method was applied to identify the essential item to compose the scales. The modelling was tested and validated using data collected from faculty members of nursing and midwifery schools in Iran.
  - Acikgul and Sad [2020] applied CVR mixed with qualitative approaches and other quantitative techniques (exploratory and confirmatory factor analysis, convergent validity, discriminant validity, nomological validity, criterion validity, internal consistency reliability, and temporal reliability) to validate the Mobile Technology Acceptance Scale for Learning Mathematics (m-TASLM) aiming to measure high school students' levels of acceptance of mobile technologies in learning mathematics.
  - Tercan and Bicakci [2023] employed Lawshe's content validity technique to validate each item that composed the perception gifted label scale (PGLS), designed to get children's perceptions about gift labels at school, which could help the design preventive educational interventions against the negative impacts of labeling.
  - In Akgul and Akturkoglu [2023], Lawshe's method was used to verify whether the efficacy of websites with specific content for facing reading and writing difficulties.
- Healthcare
  - Aiming to design and validate a yoga module for the mental health of early postpartum mothers Dogra et al. [2024], applied the CVR to a dataset composed by perceptions from thirty-eight yoga experts, to validate the essential items to compose such yoga module.
  - Findik and Aral [2023] used the content validity indices (CVI), which is based on CVR to adapt Infant-Toddler Home Observation for Measurement of the Environment (IT-HOME) to evaluate the home environment for children up to 3 years of age.
  - Hatamnejad et al. [2023] employed CVR and CVI to review and validate the questions that should remain in the translated Persian version of the Inflammatory Bowel Disease (IBD) questionnaire.

- In Khorsandi et al. [2022], it was applied CVR and CVI to content validation of PRECEDE model, a tool for measuring the knowledge, attitude, and practices of outdoor employees regarding heat stress exposure.
- Aiming to support the World Health Organization (WHO), **?** employed the CVR proposed in Lawshe [1975] to discover the essential criteria to identify and prioritize 215 health system units, according their resilient capabilities for dealing with Coronavirus, aiming to support the World Health Organization (WHO) in recognizing the capabilities of resilient health system units.

- Nutrition

  - Vergara et al. [2023] adopted the CVR to validate a survey-type questionnaire designed on knowledge, attitudes and practices of fruit and vegetable consumption in school population aged 10 to 12.
  - Murray et al. [2020] applied the Lawshe's CVR to evaluate the validity and reliability of a questionnaire measuring basic- to intermediate-level culinary skills, represented by three domains: Food Selection and Planning, Food Preparation, and Food Safety and Storage.

- Production economics and management

  - Pamucar et al. [2024] applied CVR for defining the criteria set to be used in the decision modelling for selecting the best Big Data platform able to support advanced technologies such as deep learning and machine learning.
  - Exploring the behaviors that predict success in a remote work environment, Allen et al. [2024] applied the CVI derivation of CVR to select a set of criteria classified as essential for reaching success in remote and hybrid working. This research took into account: self-evaluations from workforce, ratings provided by managers, and, comparison of results on objective performance metrics, before and after the transition to remote working.
  - Dohale et al. [2021] reported a framework to determine the compatible production system for a manufacturing firm. In this research, the CVR contributed to the process choice criteria (PCC) by measuring the consensus about which criteria should be retained as "Essential" to the modelling.
  - The approach of using CVR as a support to define the PCC by consensus measuring was also applied by Dohale et al. [2023] to support critical the decision of selecting the best-suited system from five or a hybrid (AMS + TMS) configuration: regarding the existence of four traditional manufacturing systems (TMS) (i.e. job-shop, batch-shop, mass, and continuous) and additive manufacturing system (AMS).
  - Aiming to rise the supply chain management (SCM) capacity to attract and retain workers, as a way to solve the current talent shortage in such environment, Kafa et al. [2023] used the CVR (Lawshe [1975]) to identify the essential factors that influence career advancement in SCM, taking o into account the differences in perceptions of male and female supply chain experts' about the importance of those factors.

- Quality of life

  - Upadhyay et al. [2022] utilized CvR and CvI in order to validate the development of a 20-minute yoga program with the goal of reducing burnout among healthcare worker(s).
  - In Farrell et al. [2022], the method developed by Lawshe was employed in order to compute the Content Validity Index (CVI) for two surgical skills assessment instruments: a checklist, and, a adapted form of the Objective Structured Assessment of Technical Skills (OSATS) global rating scale (GRS), surgical skills assessment instruments.
  - Claiming that there are no validated instruments for measuring the quality of life of women with long-term breast cancer in Spain, Salas et al. [2022] used CVR and CVI to validate the content of a translated version of the Long-Term Quality of Life Instrument (LTQL) for the assessment of the quality of life of long-term women breast cancer survivors.
  - Lawshe's content validity ratio (CVR) was utilized in Mukkiri et al. [2022] to examine the acceptability of each item for inclusion in the module and a scale of school preparation aimed at getting children with autism spectrum disorders (ASD) prepared for inclusive education and a scale for evaluating their school readiness.

## 3    Background on CVR concept

As the focus of this research is to improve the CVR computation, we describe in this section the original procedure proposed by Lawshe [1975] and its improvement along the time.

### 3.1 Synthesis of the Lawshe's Method

As already mentioned in the first section of this paper, Lawshe [1975] proposed the CVR technique that looks for reducing an initial set $C$ of items of a questionnaire, aiming to reduce the effort of collecting data. This method is based on performing the following steps:

1. Define an initial set $C = \{c_1, c_2, \ldots, c_m\}$ composed by $m$ criteria or variables supposed to be connected to the construct that will be under evaluation

2. Define a sample $E = \{e_1, e_2, \ldots, e_N\}$ composed by $N$ independent experts or evaluators that will evaluate

3. For each item in $C$, ask each one of the experts in $E$ to evaluate the importance of the criteria, checking one of the options in the list or scale of possibilities that is hown in Equation 1:

$$\begin{cases} \text{Essential} \\ \text{Important but not essential} \\ \text{Unnecessary} \end{cases} \tag{1}$$

4. For each item $c_k \epsilon C$, use the Equation 2 to compute the $CVR_k$ metric.

$$CVR_k = \frac{n - \frac{N}{2}}{\frac{N}{2}} \tag{2}$$

   In this equation:
   - $k$ is the index of the $k - esim$ item or variable or criteria
   - $n$ is the number of members of the sample that evaluated the item as "Essential"
   - $N$ is the number of evaluators in the sample.

5. To compare each $CVR_k$ against a critical $CVR_{min}$ defined in Lawshe [1975]. According to Lawshe [1975]:
   - If $CVR \geq CVR_{min}$, the item is classified as essential and must be retained in the questionnaire
   - Otherwise, there is no "obligation" to retain the item, and it should be discarded from the questionnaire

Table 2: A partial view of the values of $CVR_{min}$ that appears in [Lawshe, 1975]

| Number of panelists ($n_e$) | $CVR_{min}$ |
|---|---|
| 5 | 0.99 |
| 6 | 0.99 |
| 7 | 0.99 |
| 8 | 0.75 |
| 9 | 0.78 |
| ... | ... |
| 40 | 0.29 |

### 3.2 Revieweing the Lawshe's table: the proposal of Wilson et al. [2012]

According to Wilson et al. [2012], when "compared with alternative methods for quantifying content validity judgments, the *Lawshe's* method is straightforward and user-friendly, requiring only simple computations and providing a table for determining a critical cut-off value." In the other hand, Wilson et al. [2012] also mentioned some criticism to Lawshe's method, underlining the unexpected decrease in the value of $CVR_{min}$ for 8 panelists - that one can see in Table 2. As reported in Wilson et al. [2012]: "the most unsettling problem is that the statistical distribution underlying Lawshe's table is not specified."

Taking this issue into account, Wilson et al. [2012] proposed to substitute the values of $CVR_{min}$ that appear in Table 2 by a $CVR_{min}$ computed as it appears in Equation 3.

$$CVR_{min} = \frac{ne_{critical} - \frac{N}{2}}{\frac{N}{2}} \tag{3}$$

Where:

- $n$ is the number of answers collected as "Essential",
- $N$ is the number of the participants that answered the question, and,
- $ne_{critical}$ is a critical value that is a function of the sample size, the populations' mean, and the confidence level that is desirable in the modelling. According to Wilson et al. [2012], this value is computed by Equation 4:

$$ne_{critical} = \mu + Z\sigma \tag{4}$$

With:

- $\mu = N.p$ is the mean of the population,
- $N$ is the number of respondents in the sample,
- $p$ is the probability of one respondent mentions an item as "Essential",
- $\sigma = \sqrt{N.p.(1-p)}$ is the standard deviations of the answers collected, and
- $Z$ is the $z-score$ that refers to the desirable confidence level regarding a normal distribution of a sample with size equal to $N$.

As one can observe, the use of Equation 4 revised by Wilson et al. [2012] means the assumption that the distribution of probabilities should be computed through a normal approximation . The values of $ne_{critical}$ presented in Wilson et al. [2012] were computed by using a significance level $\alpha = 0.05$, which implies in: $\sum prob(n \geq ne_{critical}) \leq 0.05$.

### 3.3 The binomial validation proposed in Ayre and Scally [2014]

Ayre and Scally [2014] advise using a binomial model instead of the normal approximation suggested in Wilson et al. [2012] to determine the least number of experts needed to agree on an essential item for a particular panel size, ensuring a higher level of agreement than chance. Although Ayre and Scally [2014] recognized that there are three potential outcomes when evaluating any given item ("essential," "important, but not essential," and "not necessary"), this work assumed the outcome was dichotomous (i.e., "essential" or "not essential"). It was done because the focus of their research was in comparing their results with those of Lawshe [1975]) and Wilson et al. [2012].

As the CVR's purpose is to demonstrate a level of agreement **above** that of chance, Ayre and Scally [2014] adopted a one-tailed test with significance level ($\alpha = 0.05$) for accepting that an item is essential. Based on this reasoning, Ayre and Scally [2014] computed the minimum number of evaluators ($n_{critical}$ that are necessary to agree with assertiveness that an item is "Essential", and also recalculated such minimal number by using the CVR shown in Lawshe [1975] and also the distribution used in Wilson et al. [2012]. The proposal of Ayre and Scally [2014] should be summarized as having the following features:

- It provides the minimum number of experts that are necessary to accept an item is "Essential", with a significance level $\alpha = 0.05$, which implies in: $\sum prob(n \geq n_{critical}) \leq 0.05$
- It assumes the outcome was dichotomous (i.e., "essential" or "not essential"), despite it recognized the existence of three potential outcomes when evaluating any given item. Therefore, it assumes that the probability ($p$) of an option be randomly checked by the evaluator as $p = 0.5$.

### 3.4 Improving the computation and use of $CVR_{min}$

Despite the unquestionable contribution provided by these previous works, in this section we highlight some features we find out that should be improved in the computation and use of CVR:

1. The Equation 4 is based in the use of metric addressed to normal distributions in a situation having a scale with only three discrete (non-continous) possibilities - this criticism was highlighted and solved in Ayre and Scally [2014].

2. Although this issue has been approached by using binomial distribution as suggested in Ayre and Scally [2014], the hypothesis that $p = 1/2$ should be reviewed if the evaluator could choose one among more than two options (three options, for example, as it occurs when using the Lawshe's scale).

3. Another point to discuss is that the scale proposed in Lawshe [1975] does not cover the situation where the evaluator assumes as being not able to express his opinion about the importance of a specific item from the whole set of items. It should not be a problem in earlier researches when the data were collected by fulfilling a physical form (usually in paper media). But, in the current days it is usual to collect the data through electronic forms and, in these situations, it is also a common practice to include an option such as $NA$, that means "I do not want to answer/I can not answer/I do not know how to answer" or something like this. So, although the scale to importance remains having three points, as a matter of the facts, the user has 4 (four) options to check for each item in the form. As a consequence, a study that seeks to verify aleatory in fulfilling forms, should take into account all the four possibilities of choice in the form.

4. Suppose a case in which the members of the sample have heterogeneous perceptions about the importance of an specific item. In this case, there is the possibility of a paradoxical inconsistency: the number of respondents who agree that an item is "Essential" be sufficient to validate the item as "Essential"; and, also the number of respondents who agree that an item is "Unnecessary" be sufficient to validate the item as "Unnecessary". Therefore, there is not check for paradoxical inconsistency when using CVR.

5. Finally, we highlight that the values computed in Lawshe [1975], Wilson et al. [2012], Ayre and Scally [2014] looks to find the $n_{critical}$ in such a way that the probability of **more than** $n_{critical}$ respondents agreeing that an item is "Essential" is less than $\alpha = 0.05$. This computation does not address the discovering of the minimal (or necessary) number of respondents agreeing that an item is "Essential", to ensure that such item is actually "Essential".

## 4 Our proposal: reviewing the minimal number of experts

Our proposal is based on applying a binomial distribution suitable to the number of options the respondent has in each item of the questionnaire, and looks to discover the minimum number of experts that are necessary to validate whether an item is essential or unnecessary to be part of a questionnaire.

The Equation 6 or Equation 5 should be applied to compute the probability $prob(n)$ of $n$ members of the sample have randomly checked their answer.

$$prob(n) = \binom{N}{n} * p^n * (1-p)^{(N-n)} \tag{5}$$

or

$$prob(n) = \frac{N!}{(N-n)!n!} * p^n * (1-p)^{(N-n)} \tag{6}$$

Where:

- $N$ is the number of experts that have provided an evaluation about the relevance of the item
- $n$ is the number of experts that have randomly checked the same specific answer regarding the importance of an item
- $p$ is the probability of an evaluator have randomly assigned his answer.

Therefore, to be assigned as either "Essential" or "Unnecessary" the number of respondents that checked "Essential" or "Unnecessary" should, respectively, obey the conditions that appear in Equation 7 and Equation 8:

$$\begin{cases} n_E > \frac{N}{n_{options}} \\ n_E \geq n_{critical} \end{cases} \tag{7}$$

$$\begin{cases} n_U > \frac{N}{n_{options}} \\ n_U \geq n_{critical} \end{cases} \tag{8}$$

- $n_E$ and $n_U$ are, respectively, the number of respondents checking an item as "Essential" or "Unnecessary"
- $N$ is the size of the sample
- $n_{options}$ is the number of possibilities that the respondent could check when evaluating the relevance of an item
- $n_{critical}$ is the lowest integer number for which $prob(n_{critical}) \leq \lambda$
  - $prob(n_{critical})$ is the probability of an answer have been randomly checked by $n = n_{critical}$ members of the sample
  - $\lambda$ is a cut level that means the maximal probability that should be accepted for assuming that the answers from the sample were not aleatory. It sounds like as connected to confidence level equal to $1 - \lambda$.

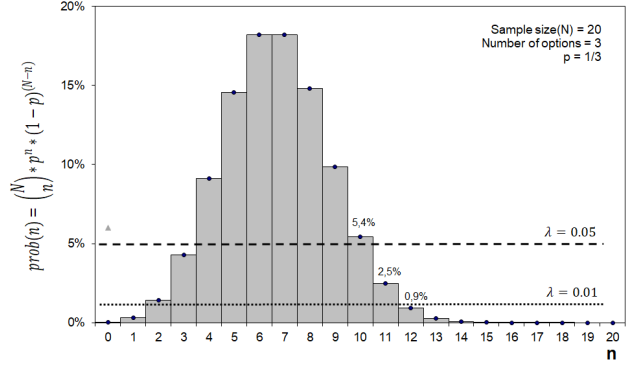

Figure 1: Enter Caption

## 4.1 An example to better describe the reasoning of our proposal

Without loss of generality, suppose a situation in which a sample composed by $N = 20$ respondents evaluates the relevance of an item to a questionnaire, having to check one of three options: "Essential","Important but not Essential", "Unnecessary". In this situation the probability $p$ of an answer be checked randomly by one respondent is $p = \frac{1}{3} = 0.33$.

In this case, one could apply Equation 9 to compute the probability $prob(n)$ of the answer have been checked aleatory by $n$ members of sample .

$$prob(n) = \binom{20}{n} * 0,33^n * (1 - 0,33)^{(20-n)} \tag{9}$$

The Figure 1 shows the distribution of probabilities for this situation. As expected, the values of the probability are smaller at the left and the right sides of the distribution.

Based on this Figure 1, the following values for $n_{critical}$ should be assigned, according the values of $\lambda$, for a sample composed by $N = 20$ members and having $p = 0.33$.

1. If $\lambda = 0.05$, then $n_{critical} = 11$
2. If $\lambda = 0.01$, then $n_{critical} = 12$

Therefore, by applying Equation 7 in the situation described in the example, we claim that to ensure that the probability of $n_E$ respondents have checked the item as "Essential" is less than $5\%$ it is necessary that the number of respondents that checked "Essential" be greater or equal than 11 respondents, that means $n_E \geq 11$. Observe that if one applies $\lambda = 0,01$, then any $n_E \geq 12$ ensures that the probability of $n_E$ have checked "Essential" is less than $1\%$ . Using an analogous reasoning, should apply Equation 8 to the example and discover that any value of $n_U \geq 12$ ensures that the probability of $n_U$ have checked "Essential" is less than $1\%$

Table 6 and Table 7 show, respectively for $p = \frac{1}{3}$ and $p = \frac{1}{4}$, the number of responses($n_{critical}$) that are needed to reject the null hypothesis (i.e: that validated the answers as non-aleatory according to the size of the sample and the cut-level level: $\lambda = 0.05$ or $\lambda = 0.01$.

## 4.2 The whole proposed algorithm

In this section we describe the whole algorithm that assigns an item of a questionnaire as either: "Essential", "Unnecessary" or "Not Essential and Unnecessary". The steps to apply this algorithm are:

1. Define an initial set $\mathbf{C} = \{c_1, c_2, \ldots, c_m\}$ composed by $m$ items candidate to be part of a questionnaire
2. Define a sample $\mathbf{E} = \{e_1, e_2, \ldots, e_N\}$ composed by $N$ independent experts or evaluators that will evaluate the relevance of the items
3. For each item in $\mathbf{C}$, ask each one of the experts in $\mathbf{E}$ to evaluate the importance of the item, using either one of the following scales:

    • $\mathbf{S_3} = \{$Essential, Important but not essential, Unnecessary$\}$, or,

- $\mathbf{S_4} = \mathbf{S_3} \bigcup \{NA\}$, where $NA$ means "I do not want to answer/I can not answer/I do not know how to answer" or something like this. In other words, each evaluator should choose an option from the set $\mathbf{S_4} = \{$Essential, Important but not essential, Unnecessary, NA$\}$

4. For each variable $c_k$, use the Equation 10 and Equation 11:

   - To compute the probability $prob(c_k)_E$ of the answer $Essential$ had been checked aleatory by $n$ members of the sample.
   - To compute the probability $prob(c_k)_U$ of the answer $Unnecessary$ had been checked aleatory by $n$ members of the sample.

$$prob(c_k)_E = \binom{N}{n_E} * p_E^n * (1-p)^{(N-n_E)} \tag{10}$$

$$prob(c_k)_U = \binom{N}{n_U} * p_U^n * (1-p)^{(N-n_U)} \tag{11}$$

In these equations:

- $N$ is the number of experts in $\mathbf{E}$ that have provided an evaluation about the relevance of the item $c_k$ as either:
  - "Essential",
  - "Essential but not necessary", or,
  - "Unnecessary".
- $n_E$ and $n_U$ are, respectively, the number of experts in $\mathbf{E}$ that have evaluated the criterion $c_k$ as:
  - $Essential$, when inferring $prob(c_k)_E$
  - $Unnecessary$, when inferring $prob(c_k)_U$
- $p$ is the probability of an evaluator alone have randomly checked his answer:

$$\begin{cases} \text{if the scale } S_3 \text{ is used, them } p = \frac{1}{3} \\ \text{otherwise, if the scale } S_4 \text{ is used, them } p = \frac{1}{4} \end{cases} \tag{12}$$

5. Define a cut level $\lambda$ according to the confidence degree desired

6. Classify whether an item must be retained or not in the questionnaire according to the decision Table 3 or, alternatively by using the decision Table 4

Table 3: Decision table based on the probabilities inferred

| | **Not** Validated as **"Unnecessary"** $prob(c_k)_U \geq \lambda$ or $n_U \leq N.p$ | Validated as **"Unnecessary"** $prob(c_k)_U \leq \lambda$ and $n_U \geq N.p$ |
|---|---|---|
| Validated as **"Essential"**: $prob(c_k)_E \leq \lambda$ and $n_E \geq N.p$ | **A: Strong recommendation to retain the item.** The item was inferred as "Essential" and not "Unnecessary" | **B: Strong paradox.** The item was paradoxically inferred as "Essential" and "Unnecessary" |
| **Not** validated as **"Essential"**: $prob(c_k)_E \geq \lambda$ or $n_E \leq N.p$ | **C. Weak paradox,** The item was not inferred as "Essential" and not "Unnecessary". So, although it is not "Essential" it is not "Unnecessary" | **D. Strong recommendation to discard the item.** The item was not inferred as "Essential" and was also inferred as "Unnecessary" |

In other words:

1. If $prob(c_k)_E \leq \lambda$ and $n \geq p * N$ and $prob(c_k)_U > \lambda$, there is an agreement that the item is essential, and, there is not an agreement that it is unnecessary. Therefore, it is the situation A, that appears in the the decision Table Table 3 where there is a strong recommendation that the item must be retained in the evaluation.

2. If $prob(c_k)_E \leq \lambda$ and $n \geq p * N$ and $prob(c_k)_U \leq \lambda$, there is a strong paradoxical incoherence or inconsistency here, once there is an agreement that the item is essential, and also an agreement that it is unnecessary. This is the situation B, and the recommendation is to proceed a deeper analysis that could include a study if the sample was really suitable for the research about this specific item.

3. If $prob(c_k)_E > \lambda$ and $prob(c_k)_U > \lambda$, there is not an agreement that the item is essential, and, also there is not an agreement that it is unnecessary. Therefore, it is another kind of paradoxical incoherence. This is the situation C, and the recommendation is to proceed a deeper analysis that could include a study if the sample was really suitable for the research about this specific item.

4. If $prob(c_k)_E > \lambda$ and $prob(c_k)_U \leq \lambda$, there is not an agreement that the item is essential, and, there is an agreement that it is unnecessary. Therefore, it is the situation D, that appears in the the decision Table Table 3 and the recommendation is that it must be eliminated from the evaluation.

Alternatively, one can use the Table 4 to decide whether an item should be or not retained in a questionnaire. The values of the variable $n_{critical}$ means the minimal number of respondents that are needed to agree that an item is essential or even if it is "not necessary".

Table 4: Decision table based on the critical number of respondents necessary to reject the hypothesis that the evaluations were aleatory

| | **Not** Validated as **"Unnecessary"** $n_U \leq n_{critical}$ | Validated as **"Unnecessary** $n_U > n_{critical}$ |
|---|---|---|
| Validated as **"Essential"**: $n_E \geq n_{critical}$ | **A: Strong recommendation to retain the item.** The item was inferred as "Essential" and not "Unnecessary" | **B: Strong paradox.** The item was paradoxically inferred as both, "Essential" and "Unnecessary" |
| **Not** validated as **"Essential"**: $n_E < n_{critical}$ | **C. Weak paradox.** The item was neither inferred as "Essential" or as "Unnecessary" | **D: Strong recommendation to discard the item.** The item was not inferred as "Essential" and was also inferred as "Unnecessary" |

Therefore:

1. If $n_E \geq n_{critical}$ and $n_E \geq p * N$ and $N_U > N.p$, there is an agreement that the item is essential, and, there is not an agreement that it is unnecessary. So, this is the situation A, that appears in the the decision Table Table 4 where there is a strong recommendation that the item must be retained in the evaluation.

2. If $N_E \geq n_{critical}$ and $n_E \geq p.N$ and $n_U \geq p.N$ $N_U \geq n_{critical}$, there is a strong paradoxical incoherence or inconsistency here, once there is an agreement that the item is essential, and also an agreement that it is unnecessary. The recommendation is to proceed a deeper analysis that could include a study if the sample was really suitable for the research about this specific item.

3. If $N_E < n_{critical}$ or $N_E < N.p$, and, $N_U \leq n_{critical}$ or $N_U < N.p$; there is not an agreement that the item is essential, and, also there is not an agreement that it is unnecessary. Therefore, this is the situation C, which is another kind of paradoxical incoherence. The recommendation is to proceed a deeper analysis that could include a study if the sample was really suitable for the research about this specific item.

4. If $N_E < n_{critical}$ or $N_E < N.p$, and, $N_U \geq n_{critical}$ and $N_U \geq < N.p$; there is not an agreement that the item is essential, and, there is an agreement that it is unnecessary. Therefore, it is the situation D, that appears in Table 4 the recommendation to dropout the item from the questionnaire.

Aiming to facilitate the use of $n_{critical}$ Table 6 and Table 7 in Appendix A illustrate, respectively for $S_3$ and $S_4$, the number of responses that are needed ($n_{critical}$) to reject the null hypothesis (i.e: that validated the answers as non-aleatory).

The values in these tables cover a size sample from $N = 5$ up to $N = 100$, and the cut-levels: $\lambda = 0.05$ or $\lambda = 0.01$). An extended version of these tables, covering samples with size up to $10,000$ and the same cut-levels, is available in a open repository and will be mentioned here after paper approval.

## 5 Comparing our proposal against the previous ones

Table 8, that appears in Appendix A shows from its second column up to the fifth one the values of $n_{critical}$ regarding the sample size that appears in the first column, and computed by using our proposal with cut-level ($\lambda$) equal to $0.05$ and $0.01$, respectively. The last to columns shows the values of CVR computed as proposed in Ayre and Scally [2014]

and in Wilson et al. [2012], respectively. As one can see, there are differences among the values, which increases as the size of the sample grows. Another point is that the values of $n_{critical}$ for a cut-level $\lambda = 0.05$ are lower than those computed by Wilson et al. [2012] and Ayre and Scally [2014] with a confidence degree $\alpha = 0.05$, which means that some items do not classified by the previous works as "Essential" should be assigned as "Essential" by the Binomial Cut-level Validity (BCV).

In this section we also compare the features of BCV with those that are present in the previous works related to CVR. Table 5 summarizes the results of this comparison.

Table 5: Comparing the features of BCV with the previous works

|  | Lawshe | Wilson | Ayres | BCV (Our proposal) |
| --- | --- | --- | --- | --- |
| Distribution | Not declared | A normal aproximation of the Binomial | Binomial Distribution | Binomial Cut-level |
| Scale size | Three points | Three points | Three points | Three or four points |
| Probability of a respondent choose an option from the scale | 50% | 50% | 50% | 33.3% for the scale with 3 points, or, 25% for the scale with 4 points |
| Validation if an item is "Essential" | Yes | Yes | Yes | Yes |
| Validation if an item is "Not necessary" | No | No | No | Yes |
| Paradox checking | No | No | No | Yes |
| Validation as non aleatory | Not mentioned | Significance level $\alpha = 0.05$ | Significance level $\alpha = 0.05$ | Cut level: $\eta = 0.05$ and $\tau = 0.01$ |

Observing , we infer the main improvements provided by our proposal are:

- the possibility to verify it an item is recommended to be discarded,

- its power to identify the presence of incoherence or paradox issues in the survey, and,

- the use of cut-values ($\lambda$) suitable to the binomial probability function and the number of options in the scale of the survey.

## 6 Conclusion

The Content Validity Ratio (CVR), as described in Lawshe [1975], has been recognized as a prominent and valid tool for measuring the importance of a questionnaire item. However, by analyzing the original proposal and its main variants, we concluded there are some points to be improved regarding in the computation and use of CVR:

- When employing CVR, no evaluation is performed to detect the presence of conflicting inconsistency. In essence, it just checks whether the number of replies in the sample is sufficient to classify an item as "Essential". Nonetheless, a sample may paradoxically state an item as both "essential" and "unnecessary". While this particular contradiction is relatively rare, it makes sense to check if it happens.

- Since each respondent is required to select one choice from a set of three options ("Essential", "Important, but not Essential", and "Unnecessary"), this circumstance can be seen as a binomial situation with a probability of success (p) equal to $p = \frac{1}{3}$. However, the CVR is calculated under the assumption that $p = \frac{1}{2}$ and employs a normal approximation rather than a binomial one.

    • In a binomial situation with three options ("Essential", "Important, but not Essential", and "Unnecessary"), the chance of a sample member selecting one of these options at random is p = 1/3. However, the CVR is calculated with the assumption that p = 1/2 and using a normal approximation rather than a binomial one.

- Since the binomial probability distribution, which depends on the number of respondents, is not continuous, to rely on approximations that are designed for continuous functions is less precise than directly using the binomial probability function.

The Binomial Cut-Level Validity (BCV) proposed here addresses these criticisms by presenting a strategy that enables:

- A "double check validation" test is provided to detect paradoxical contradictions in the results. This test examines if an item has been classified as "Essential" or "Unnecessary".

- The use of a suitable binomial distribution assuming $p$ as a function of the number of options the evaluator has, instead of using a normal approximation and assuming $p = \frac{1}{2}$. This makes possible to use the proposal with scales with a number of options different than three.

- The adoption of a cut-level technique that is based on the computation of probability as a binomial function instead of using an approximation which is based on integrating a non continuous function.

In other words our study proposes the Binomial Cut-level Validity (BCV) as an alternative metric to the CVR, providing a confident and pioneer whole classification of the potential items into one of the following categories:

- Essential

- Not necessary

- Paradoxically inconsistent type A (classified by the sample as "Essential" and also "Unnecessary")

- Paradoxically inconsistent type B (classified by the sample as not "Essential" and also as not "Unnecessary")

We also consider as a enhancement delivered by this work the utilization of distinct cut levels $\lambda$ (set at 0.01 or 0.05), which facilitated the reassessment and expansion of understanding on the minimum number of respondents required to reach a consensus on the importance of an item.

The results presented in Table 8 indicate that BCV is more cautious in its classifications, as it is capable of identifying "Essential" items that are not identified as such by CVR and its variations.

As mentioned in the previous section, when compared against previous works, our proposal is innovative by introducing the verification if an item is recommended to be discarded and the existence of any paradoxical incoherence in the data collected in the survey. It also introduces cut-levers more suitable to the data distribution.An item is classified as 'Paradoxically inconsistent' if the evaluations collected from the survey results in classifying an item as both: Essential and Not Necessary, which is an incoherence that means that the sample has heterogeneous perception about the need of an item.

**Further works**

As further researches we suggest to conduct a comprehensive experiment encompassing a wide array of situations to evaluate the sensitivity of results based on comparing BCV (present study), Lawshe (Lawshe [1975]), Wilson (Wilson et al. [2012]), and Ayre (Ayre and Scally [2014]). We also propose a similar approach to investigate the effects of Using BCV instead of CVR as an input to the computation of CVI.

Finally, we plan to carry out additional investigation regarding the adaptation our proposal for supporting multi-criteria decision and negotiation modeling. It makes sense once identifying the essential criteria is this subject is actually a complex task surrounded by conflicting objectives and interests. This should be a wide and useful potential contribution of this work to this another field.

# ACKNOWLEDGEMENTS

This study was financed in part by Coordenação de Aperfeiçoamento de Pessoal de Nível Superior (CAPES), Grant/Award Number: 001; Conselho Nacional de Desenvolvimento Científico e Tecnológico (CNPQ), Grant/Award Number: 314953/2021-3 and 421779/2021-7; and, Fundação Carlos Chagas de Amparo a Pesquisa do Estado do Rio de Janeiro (FAPERJ), Grant/Award Number: 200.974/2022.

Will be provided after paper acceptance.

# CONFLICT OF INTEREST

The authors declare no conflicts of interest.

# DATA AVAILABILITY STATEMENT

A expanded version of the data that appears in in Table 6 and Table 7 is available in Gomes Costa [2023], with free access and use only constrained by the need of including a citation to this work.

## A Appendix

Table 6: Values of $n_{critical}$ for a scale $S_3$ (with three options), according to the sample size ($N$) and the cut-level ($\lambda = 0.05$ and $\lambda = 0.01$)

| N | $\lambda = 0.05$ | $\lambda = 0.01$ | N | $\lambda = 0.05$ | $\lambda = 0.01$ | N | $\lambda = 0.05$ | $\lambda = 0.01$ |
|---|---|---|---|---|---|---|---|---|
| 5 | 5 | 5 | 37 | 17 | 20 | 69 | 28 | 32 |
| 6 | 5 | 6 | 38 | 17 | 20 | 70 | 28 | 32 |
| 7 | 5 | 6 | 39 | 18 | 20 | 71 | 29 | 33 |
| 8 | 6 | 7 | 40 | 18 | 21 | 72 | 29 | 33 |
| 9 | 6 | 7 | 41 | 18 | 21 | 73 | 29 | 34 |
| 10 | 7 | 8 | 42 | 19 | 22 | 74 | 30 | 34 |
| 11 | 7 | 8 | 43 | 19 | 22 | 75 | 30 | 34 |
| 12 | 7 | 9 | 44 | 19 | 22 | 76 | 30 | 35 |
| 13 | 8 | 9 | 45 | 20 | 23 | 77 | 31 | 35 |
| 14 | 8 | 10 | 46 | 20 | 23 | 78 | 31 | 35 |
| 15 | 9 | 10 | 47 | 20 | 24 | 79 | 32 | 36 |
| 16 | 9 | 11 | 48 | 21 | 24 | 80 | 32 | 36 |
| 17 | 9 | 11 | 49 | 21 | 24 | 81 | 32 | 37 |
| 18 | 10 | 12 | 50 | 22 | 25 | 82 | 33 | 37 |
| 19 | 10 | 12 | 51 | 22 | 25 | 83 | 33 | 37 |
| 20 | 11 | 12 | 52 | 22 | 25 | 84 | 33 | 38 |
| 21 | 11 | 13 | 53 | 23 | 26 | 85 | 34 | 38 |
| 22 | 11 | 13 | 54 | 23 | 26 | 86 | 34 | 38 |
| 23 | 12 | 14 | 55 | 23 | 27 | 87 | 34 | 39 |
| 24 | 12 | 14 | 56 | 24 | 27 | 88 | 35 | 39 |
| 25 | 13 | 15 | 57 | 24 | 27 | 89 | 35 | 40 |
| 26 | 13 | 15 | 58 | 24 | 28 | 90 | 35 | 40 |
| 27 | 13 | 15 | 59 | 25 | 28 | 91 | 36 | 40 |
| 28 | 14 | 16 | 60 | 25 | 29 | 92 | 36 | 41 |
| 29 | 14 | 16 | 61 | 25 | 29 | 93 | 36 | 41 |
| 30 | 14 | 17 | 62 | 26 | 29 | 94 | 37 | 41 |
| 31 | 15 | 17 | 63 | 26 | 30 | 95 | 37 | 42 |
| 32 | 15 | 17 | 64 | 26 | 30 | 96 | 37 | 42 |
| 33 | 15 | 18 | 65 | 27 | 31 | 97 | 38 | 43 |
| 34 | 16 | 18 | 66 | 27 | 31 | 98 | 38 | 43 |
| 35 | 16 | 19 | 67 | 27 | 31 | 99 | 38 | 43 |
| 36 | 17 | 19 | 68 | 28 | 32 | 100 | 39 | 44 |

Table 7: Values of $n_{critical}$ for a scale $S_4$ (with three options), according to the sample size ($N$) and the cut level ($\lambda = 0.05$ and $\lambda = 0.01$)

| N | $\lambda = 0.05$ | $\lambda = 0.01$ | N | $\lambda = 0.05$ | $\lambda = 0.01$ | N | $\lambda = 0.05$ | $\lambda = 0.01$ |
|---|---|---|---|---|---|---|---|---|
| 5 | 5 | 5 | 37 | 14 | 16 | 69 | 22 | 26 |
| 6 | 5 | 5 | 38 | 14 | 16 | 70 | 22 | 26 |
| 7 | 5 | 6 | 39 | 14 | 17 | 71 | 23 | 26 |
| 8 | 5 | 6 | 40 | 14 | 17 | 72 | 23 | 27 |
| 9 | 5 | 6 | 41 | 15 | 17 | 73 | 23 | 27 |
| 10 | 6 | 7 | 42 | 15 | 18 | 74 | 23 | 27 |
| 11 | 6 | 7 | 43 | 15 | 18 | 75 | 24 | 28 |
| 12 | 6 | 8 | 44 | 16 | 18 | 76 | 24 | 28 |
| 13 | 7 | 8 | 45 | 16 | 19 | 77 | 24 | 28 |
| 14 | 7 | 8 | 46 | 16 | 19 | 78 | 25 | 28 |
| 15 | 7 | 9 | 47 | 16 | 19 | 79 | 25 | 29 |
| 16 | 8 | 9 | 48 | 17 | 19 | 80 | 25 | 29 |
| 17 | 8 | 9 | 49 | 17 | 20 | 81 | 25 | 29 |
| 18 | 8 | 10 | 50 | 17 | 20 | 82 | 26 | 30 |
| 19 | 8 | 10 | 51 | 17 | 20 | 83 | 26 | 30 |
| 20 | 9 | 10 | 52 | 18 | 21 | 84 | 26 | 30 |
| 21 | 9 | 11 | 53 | 18 | 21 | 85 | 26 | 30 |
| 22 | 9 | 11 | 54 | 18 | 21 | 86 | 27 | 31 |
| 23 | 10 | 12 | 55 | 18 | 22 | 87 | 27 | 31 |
| 24 | 10 | 12 | 56 | 19 | 22 | 88 | 27 | 31 |
| 25 | 10 | 12 | 57 | 19 | 22 | 89 | 27 | 32 |
| 26 | 11 | 13 | 58 | 19 | 23 | 90 | 28 | 32 |
| 27 | 11 | 13 | 59 | 20 | 23 | 91 | 28 | 32 |
| 28 | 11 | 13 | 60 | 20 | 23 | 92 | 28 | 32 |
| 29 | 11 | 14 | 61 | 20 | 23 | 93 | 28 | 33 |
| 30 | 12 | 14 | 62 | 20 | 24 | 94 | 29 | 33 |
| 31 | 12 | 14 | 63 | 21 | 24 | 95 | 29 | 33 |
| 32 | 12 | 14 | 64 | 21 | 24 | 96 | 29 | 34 |
| 33 | 13 | 15 | 65 | 21 | 25 | 97 | 29 | 34 |
| 34 | 13 | 15 | 66 | 21 | 25 | 98 | 30 | 34 |
| 35 | 13 | 15 | 67 | 22 | 25 | 99 | 30 | 34 |
| 36 | 13 | 16 | 68 | 22 | 25 | 100 | 30 | 35 |

Table 8: Values of $n_{critical}$ computed by using BCV and by using the previous works based on CVR

| | Binomial Cut-Level Validity, BCV (This article) | | | | Wilson et al. (2012) | Ayre and Scally (2014) |
| | Scale=3 options Binomial Distribution | | Scale=4 options Binomial Distribution | | Scale=3 options Normal Approximation | Scale=3 options Binomial Distribution |
| N | $p = 1/3$ $\lambda$=.05 | $p = 1/3$ $\lambda$=.01 | $p = 1/4$ $\lambda$=.05 | $p = 1/4$ $\lambda$=.01 | $p = 1/2$ $\alpha = .05$ | $p = 1/2$ $\alpha = .01$ |
|---|---|---|---|---|---|---|
| 5 | 5 | 5 | 5 | 5 | 4 | 5 |
| 6 | 5 | 6 | 5 | 5 | 5 | 6 |
| 7 | 5 | 6 | 5 | 6 | 6 | 7 |
| 8 | 6 | 7 | 5 | 6 | 6 | 7 |
| 9 | 6 | 7 | 5 | 6 | 7 | 8 |
| 10 | 7 | 8 | 6 | 7 | 8 | 9 |
| 11 | 7 | 8 | 6 | 7 | 8 | 9 |
| 12 | 7 | 9 | 6 | 8 | 9 | 10 |
| 13 | 8 | 9 | 7 | 8 | 9 | 10 |
| 14 | 8 | 10 | 7 | 8 | 10 | 11 |
| 15 | 9 | 10 | 7 | 9 | 11 | 12 |
| 16 | 9 | 11 | 8 | 9 | 11 | 12 |
| 17 | 9 | 11 | 8 | 9 | 12 | 13 |
| 18 | 10 | 12 | 8 | 10 | 12 | 13 |
| 19 | 10 | 12 | 8 | 10 | 13 | 14 |
| 20 | 11 | 12 | 9 | 10 | 14 | 15 |
| 21 | 11 | 13 | 9 | 11 | 14 | 15 |
| 22 | 11 | 13 | 9 | 11 | 15 | 16 |
| 23 | 12 | 14 | 10 | 12 | 15 | 16 |
| 24 | 12 | 14 | 10 | 12 | 16 | 17 |
| 25 | 13 | 15 | 10 | 12 | 17 | 18 |
| 26 | 13 | 15 | 11 | 13 | 17 | 18 |
| 27 | 13 | 15 | 11 | 13 | 18 | 19 |
| 28 | 14 | 16 | 11 | 13 | 18 | 19 |
| 29 | 14 | 16 | 11 | 14 | 19 | 20 |
| 30 | 14 | 17 | 12 | 14 | 19 | 20 |
| 31 | 15 | 17 | 12 | 14 | 20 | 21 |
| 32 | 15 | 17 | 12 | 14 | 21 | 22 |
| 33 | 15 | 18 | 13 | 15 | 21 | 22 |
| 34 | 16 | 18 | 13 | 15 | 22 | 23 |
| 35 | 16 | 19 | 13 | 15 | 22 | 23 |
| 36 | 17 | 19 | 13 | 16 | 23 | 24 |
| 37 | 17 | 20 | 14 | 16 | 23 | 24 |
| 38 | 17 | 20 | 14 | 16 | 24 | 25 |
| 39 | 18 | 20 | 14 | 17 | 25 | 26 |
| 40 | 18 | 21 | 14 | 17 | 25 | 26 |